\documentstyle[twocolumn,prb,aps,epsf]{revtex}

\begin{document}
\twocolumn[
\title{A strong-coupling theory of superfluid ${^4}$He.}
\author{N. Gov $^\dagger$ and E. Akkermans$^{\dagger *}$}

\address{$^\dagger$ Department of Physics, Technion, 32000 Haifa, Israel \\
$^*$ Physique des Solides et LPTMS, Universite` Paris-Sud 91405, Orsay,Cedex,
France.}

\maketitle
\tightenlines
\widetext
\advance\leftskip by 57pt
\advance\rightskip by 57pt

\begin{abstract} 
We propose a new theoretical approach to
the excitation spectrum of superfluid ${^4}He$. 
It is based on the assumption that, in addition to the usual Feynman 
density fluctuations, there exist localized modes which describe the 
short range behaviour in the liquid associated with microscopic cores of quantized vortices.
We describe in a phenomenological way the hybridization of 
those two kinds of excitations and we compare the resulting energy spectrum 
with experimental data, e.g. the structure factor and the cross section for single quasi-particle excitations.
 We also predict the existence of another type of 
excitation interpreted as a vortex loop. The energy of this mode agrees both with critical velocity experiments 
and high energy neutron scattering.
In addition we derive a relation between the condensate fraction and the roton energy and we calculate 
the reduction of the ground-state energy due to the superfluid order. 
\end{abstract} 

\vskip 0.2cm
]

\narrowtext
\tightenlines
 
\vspace{.2cm}

\section{Introduction} 
 
In this paper we present an alternative description of the energy spectrum of superfluid $^4$He.
 Among the standard approaches to this problem, we mainly find either the effective Hamiltonian of Bogoliubov \cite{bogo}, 
or the variational approach proposed by Feynman \cite{feynman}.

The Bogoliubov description relies on the assumption of a weakly interacting Bose gas. 
Its main success was to obtain a linear dispersion for the long wavelength excitations 
starting from the quadratic spectrum of free bosons. This linear phonon branch cannot be obtained from simple 
perturbation theory but corresponds to a RPA-like description of this problem. This linear behavior
 is indeed one of the main features of the experimental spectrum (fig.1).
Nonetheless this success, neither the initial slope of the dispersion $\varepsilon (k)$, nor the rest of the spectrum beyond
 the linear part, is quantitatively obtained from this approach.

A alternative description was proposed by Feynman \cite{feynman}, based on a variational form of the wavefunction for
 the low-lying excited states which leads to an excitation spectrum of the form
\begin{equation}
\varepsilon (k)=\frac{\hbar ^2k^2}{2mS(k)}
\label{feynspec}
\end{equation}
where $m$ is the mass of the atoms and $S(k)$ is the static structure factor of the liquid.
This expression, using the experimental structure factor \cite{cowley}, is shown in fig.1. It reproduces 
quantitatively the low momentum linear behavior, since in this limit the f-sum rule is exhausted 
by these density fluctuations \cite{pines}. Unlike the weakly interacting Bose gas limit, it also reproduces qualitatively
 the roton minimum.
Despite these successes, this variational wavefunction, although it includes Bose symmetry, does not contain special
 features characterizing the superfluid phase. For instance, since it contains only information about the static structure
 factor, it gives essentially the same energy spectrum (\ref{feynspec}) both below $T_{\lambda}$ in the superfluid phase
 and in the normal phase ($T>T_{\lambda}$) where experimental results exhibit very different behaviors for the spectrum 
of these two phases. Unlike $S(k)$ which almost does not change, the 
dynamical structure factor $S(k,\omega)$ is very sensitive to the superfluid transition. This shows up in the vanishing of
 the roton and maxon peaks above $T_{\lambda}$ \cite{griffin}.

In order to get a better quantitative agreement in the roton region, subsequent refinements \cite{feyncohen,reatto} 
of the original variational wavefunction were proposed, which include more localized structure on top of the 
delocalized density fluctuations (phonons). This was achieved by introducing more variational parameters, at the expenses
 of the initial interpretation of the excitations as density fluctuations. Despite sizeable improvements
 for the energy spectrum and other measurable quantities including the scattering intensity, this approach does not fully 
answer the question of the nature of the superfluid behavior.

An alternative approach to the superfluididty of $^4$He using path integral Monte Carlo 
 calculations \cite{ceperley} was proposed. Although it does not provide explicit results for the spectrum, 
it showed unambiguosly how
 superfluidity is related to the coherent exchange of atoms along rings of all sizes.

Our motivation in order to build a phenomenological model for the superfluid is based on the features discussed previously.
 We assume that the main features of superfluidity show up in the spectrum in the large momentum  
range above the linear phonon part. This includes the roton minimum (at energy $\Delta$) and the saturation for very
 large $k$'s at the energy $2\Delta$ \cite{pisto}. We propose to view superfluid $^4$He as 
 the condensation in one state of a set of localized states 
with long-range interactions. This is in contrast to the model of hard-core bosons on a lattice without
 interactions. 
We shall describe these local modes as two-level systems (TLS) of bare energy $E_0$. This approach is analogous to the 
description of dipolar glasses \cite{glass} where such a picture, based on two-level tunneling states,
 has been successfully used to describe the low temperature excitations measured by specific heat or heat transport experiments \cite{glassx}.

The TLS we consider are localized on atomic scales and may correspond to
quantum coherent rings of a few atoms only, which
 are supposed to be responsible for the superfluid behaviour of the liquid.
These short rings correspond to vortex cores of sizes $R \simeq 1 \AA$, and may be considered as local dipoles. 
Large ring structures arise from the long-range interaction between the TLS.

The effective Hamiltonian $H_{loc}$ we propose for these interacting local modes can be diagonalized by a Bogoliubov
 transformation. The resulting spectrum bears similarity with the weakly interacting Bose gas expression, although it
 does not assume a weak interaction. Indeed, in order to recover a gapless phonon branch at low momentum ($k\simeq 0$), 
we need to consider strong interactions, namely interactions of the order of the bare energy $E_0$.
Unlike ordered structures of dipoles on a lattice for which it is possible to calculate the interaction term, for
 liquid $^4$He (and for glasses), the interaction term is obtained from the experimental energy spectrum.
 Despite this shortcoming, this model gives closed analytical expressions for quantities such as the scattering 
intensity of neutrons by single quasiparticle excitations, the relative variation of the ground-state energy and 
the condensate fraction in terms of a single quantity, namely the strength of the interaction, so that 
it becomes possible to relate them.

In order to compare this approach to the Feynman variational scheme, we use an equivalent formulation of our problem 
based on a description of the interaction between the localized modes as due to a virtual exchange of Feynman phonons.
 This is very much in the way Hopfield \cite{hopfield} and Anderson
\cite{anderson} considered the exciton problem in dielectric media. This description
 is equivalent to the previous one based on the effective Hamiltonian $H_{loc}$ provided we use
 the dipolar approximation which corresponds to large momentum. For low momentum, both the
 density fluctuations and the large scale exchange cycles correspond to the same
 degrees of freedom and cannot be disentangled anymore. 
In the large momentum limit the energy spectrum results from the hybridization of these two kinds of excitations, which
 represent independent degrees of freedom of the superfluid. This way we gain a better understanding of the roton
 part of the spectrum.
In addition we find a new localized excitation branch at energy $E_{2}=2E_0$, which we interpret as 
microscopic vortex-loop excitations of the superfluid. These type of excitations are usually thought as being 
completely independent of the phonon-roton excitations, while here we propose a model that describes them both together. 

This paper is organized as follows.
We begin by introducing the effective Hamiltonian of the strong coupling 
description in section II. In section III, we obtain expressions for the 
scattering intensity of the single-excitation branch and compare it to the experimental 
data. In section IV, we evaluate the reduction in the ground-state energy of 
the superfluid and the condensate fraction. In 
section V, we consider the dipolar approximation and the resulting 
hybridization scheme. We compare the resulting spectrum and 
the static structure factor to the experimental data. Finally, in section VI, we discuss 
the vortex-loop branch and we conclude in section VII. 
 
\section{The effective Hamiltonian description} 
 
We consider superfluid $^4$He as a set of local states described as two-level systems of bare energy $E_0$. Their
 Hamiltonian is then
\begin{equation} 
{H^0_{loc}}={\sum_k}{E_0}{{b_k}^{\dagger }}{b_k}
\label{hloc0} 
\end{equation} 
 where translational invariance allows us to write operators in $k$-space. The operators $b_k$ obey bosonic
 commutation relations. This holds in the 
limit of a low density of localized modes ~\cite{anderson}, i.e. for the case of no multiple occupation of a site.
 This condition corresponds to vortex-cores or small exchange rings which cannot be multiply excited on the same site.
 The value of $E_0$ will be determined later on.
 
The interaction between these localized-modes can generally 
be written as~\cite{anderson}  
\begin{eqnarray} 
H_{loc} &=&{\sum_k}\big({E_0}+X(k)\big){{b_k}^{\dagger }}{b_k}  \label{hloc} \\
&&\ \ +{\sum_k}X(k)\left( {{b_k}^{\dagger }}{b_{-k}^{\dagger }}+b_kb_{-k}\right)  \nonumber
\end{eqnarray} 
The diagonal part in this Hamiltonian corresponds to the hopping of a local mode between sites,
 and the off-diagonal part describes the creation (or the anihilation) of pairs of local modes on distinct sites.
The matrix element $X(k)$ depends on the microscopic details of the interaction.
For a lattice structure like excitons in a crystal \cite{heller} or solid \cite{bcc} bcc $^4$He,
it is possible to express $X(k)$ as a summation of dipolar 
terms. This is not possible in the superfluid phase, for
which we have no lattice structure. 
 
The Hamiltonian $H_{loc}$ is diagonalized by the Bogoliubov 
transformation ${\beta _k}=u(k)b_k+v(k)b{^{\dagger }}_{-k}$. The 
resulting spectrum is 
\begin{equation} 
E(k)=\sqrt{{E_0}({{E_0}+2X(k)})}  \label{ek} 
\end{equation} 
 and the two functions $u(k)$ and $v(k)$ are given by  
\begin{eqnarray} 
{u^2}(k)&=&{\frac 12}\left( {\frac{{{E_0}+X(k)}}{{E(k)}}}+1\right)  \nonumber \\
{v^2}(k)&=&{\frac 12}\left( {\frac{{{E_0}+X(k)}}{{E(k)}}}-1\right) 
\label{uv} 
\end{eqnarray} 
 The corresponding ground-state wavefunction for the local modes is \cite{huang}  
\begin{equation} 
\left| \Psi _0\right\rangle =\prod_k\exp \left( \frac{v_k}{u_k}{{b_k}%
^{\dagger }}{b_{-k}^{\dagger }}\right) \left| vac\right\rangle  \label{psi0} 
\end{equation} 
where $\left| vac\right\rangle$ is the state with no local modes, i.e. anihilated by
 $b_{k}$, namely $b_{k} \left| vac\right\rangle =0$.
The ground-state is a coherent state of pairs of 
localized-modes. This coherent state defines a
global phase which, once fixed, breaks the gauge-invariance.
The excitation given by $\beta {^{\dagger }}_{k} \left| \Psi _0\right\rangle$ corresponds to the breaking
 of a pair of momentum $k$, leaving an unpaired local-mode.

The expressions of the bare energy $E_0$ and $X(k)$ stem from the following remarks. The equality $E(k)=E_0$
 is achieved in the zero coupling limit ($X(k)=0$) and it corresponds to a normal phase for the local modes  
characterized by $v_{k}=0$ in the relations (\ref{uv}) and (\ref{psi0}). This corresponds to the termination point of
 the energy spectrum \cite{pisto} for which, at large momentum $k$, the energy goes to $2\Delta$ where $\Delta$ is 
the roton energy (fig.1).
Beyond the termination point, there is no quasi-particle excitation and only a structureless free-recoil scattering
 characteristic of a normal liquid.
We shall therefore take for the bare energy the value $E_{0}=2 \Delta $ \cite{note}.

In the opposite limit, i.e. for low momentum $k \rightarrow 0$, we demand that the spectrum (\ref{ek}) will be gapless,
 namely $E(0)=0$. This requires
\begin{equation} 
X(0)={E_{0} \over 2}= -\Delta  
\label{xk0} 
\end{equation} 
 
Notice that for dipoles on a lattice \cite{bcc}, this condition
is a self-consistent
definition of $E_0$ since the energy required to excite
locally a dipole from the ground-state corresponds to the one needed to flip it with respect to
the coherent background. This excitation changes the sign of the interaction energy and costs 
$2\mid{X(k=0)}\mid=2\Delta=E_0$, which corresponds to the previous definition of the energy of the local mode.
 
For arbitrary $k$, $X(k)$ is obtained by comparing Eq.(\ref{ek}) with the measured energy spectrum (fig.2).
This shape of $X(k)$ is similar to the expression calculated for dipoles on a simple 
cubic lattice along various directions \cite{heller,bcc}. 
This could be helpful towards the development of a microscopic model for the superfluid phase.
 
It is useful at this stage to compare our approach with the weakly interacting Bose gas limit. There, the
 Hamiltonian is given within the Bogoliubov approximation by
\begin{eqnarray} 
H_{WIBG}&=&\sum_{k}\left( \varepsilon _k+N_0V_k\right) a_k^{\dagger}a_k \nonumber \\
&+&\sum_{k}N_0V_k\left( a_k^{\dagger }a_{-k}^{\dagger }+a_ka_{-k}\right) 
\label{wibg} 
\end{eqnarray} 
 where $N_0=\left| \left\langle a_0\right\rangle \right| ^2$, $%
\varepsilon _k=\hbar ^2k^2/2m$, and $V_k$ is the effective potential between 
the bosons at wave vector $k$. The operators $a_k^{\dagger }$ and $a_k$ correspond
to the creation and anihilation of an atom and not of a localized-mode as in (\ref 
{hloc}). 
The Bogoliubov excitation spectrum which corresponds to (\ref{wibg}) is 
\begin{equation} 
E=\sqrt{\varepsilon _k\left( \varepsilon _k+2N_0V_k\right) }
\label{ebogo} 
\end{equation} 
This spectrum is linear in the $k\rightarrow 0$ limit ( $E \simeq \hbar kc$),
with a sound velocity defined by $c=\sqrt{N_0V_0/m}$. 

The physics described by the two Hamiltonians $H_{loc}$ and $H_{WIBG}$ is very 
different. The weakly interacting Bose gas has a non-interacting limit given by the ideal Bose gas 
which, at T=0, is known to be fully condensed in the state $k=0$ and  which has already a broken 
gauge symmetry. The interaction depletes the condensate and changes the spectrum to linear at low momentum.
 On the other hand, the Hamiltonian $H_{loc}$ corresponds to a set of 
independent localized modes with no broken symmetry in the non-interacting limit ($X(k)=0$). 
To restore in the linear momentum regime both the condensation and broken gauge symmetry,
we need a non-zero 
interaction $X(k)$. Since this interaction is large ($X(0)=-E_{0}/2$), it is not a small perturbation to 
the non-interacting case. In that sense, it is a strong-coupling description of 
the superfluid. 
 
We would like to point out that the entire excitation spectrum given by
(\ref{ek}) is unique to the superfluid phase, as only there we can
define the function $X(k)$ and local mode energy $E_0$. In particular, the phonon
mode is different from the zero-sound mode that appears in the normal
fluid \cite{nepo}.
We shall now use this approach to obtain analytical results that we shall 
compare subsequently with experimental measurements.
 
\section{Scattering intensity} 
 
The neutron scattering intensity is a direct probe of the density 
fluctuations in the liquid and may be described using the dynamic 
structure factor $S(k,\omega )$. It is usually accepted since the work of 
Miller, Pines and Nozieres ~\cite{mpn} that we can split the total 
contribution to $S(k,\omega )$ into two parts,  
\begin{equation} 
S(k,\omega )=NZ(k)\delta (\hbar \omega -{\varepsilon (k)})+{S^{(1)}}(k,\omega ) 
\label{zpines} 
\end{equation} 
where the first term accounts for single quasi-particle excitations of weight $Z(k)$, while 
the second describes multiparticle excitations. This separation is
justified at low temperature and low momentum, typically for $k\leq 0.5%
{\AA }^{-1}$. In this regime, integrating (\ref{zpines}) over the energy and 
noticing that ${S^{(1)}}(k,\omega )$ vanishes in the low momentum limit we 
obtain $Z(k)=S(k)$, which results also from the Feynman theory. The 
comparison with the experimental data ~\cite{cowley} shows that it 
works indeed in the low momentum regime mentionned above, but fails to describe the 
non linear part of the spectrum, except perhaps for the position of the 
maximum. 
 
Since the excitations of the local modes from the ground state (\ref{psi0}) involve breaking pairs of 
coherent local modes,
the probability of a neutron to scatter
inelastically on such an excitation is proportional to the occupation density of
these pairs at each momentum given by
$\langle {{b_k}^{\dagger }}{b_{-k}^{\dagger }}\rangle ={u_k}{v_k}$, so that
\begin{equation}
Z(k)=4\pi {k^2}{I_0}{u_k}{v_k}  \label{zk0}
\end{equation} 
where $I_0$ is an arbitrary normalization constant and where the factor $4\pi {k^2}$ comes from 
the three dimensional phase space.
 Using the expression (\ref{uv}) for $%
u_k$ and $v_k$, we obtain ${u_k}{v_k}={\frac 12}{\frac{{|X(k)|}}{{%
E(k)}}}$ which together with (\ref{ek}) gives   
\begin{equation} 
Z(k)=\pi {k^2}{I_0}{\frac{{{E_0}}}{{E(k)}}}\left| \left( \frac{%
E(k)}{E_0}\right) ^2-1\right|  \label{zk} 
\end{equation} 
Using the experimental $E(k)$ in expression (\ref{zk}) we obtain a 
differential cross-section which agrees well with the experimental results \cite{cowley}) obtained at
saturated vapor pressure (S.V.P.) as shown 
in fig.3. In the low momentum limit, we recover the 
proportionality between $S(k)$ and $Z(k)$ and therefore $Z(k)$ is linear with $k$.
We emphasize that as a result of both (\ref{zk}) and the saturation of $E(k)$ to $2\Delta$ at large $k$, 
the scattering intensity $Z(k)$ vanishes identically at this point. This indeed corresponds to the experimental
 measurement of $Z(k)$, and could not be obtained neither from weakly interacting Bose gas description nor by 
variational approaches.
 
\section{Description of the condensate} 
 
\subsection{Condensate fraction}

In the microscopic description of a Bose liquid, the condensate is usually characterized by the condensate fraction,
 $n_{0}/n$, which measures the relative occupation by the atoms of the lowest energy state. At zero temperature it 
is equal to one for an ideal Bose gas, while a finite interaction depletes the lowest energy state with a corresponding
 decrease of $n_{0}/n$.
It is possible to calculate this fraction in terms of the interaction potential only within the Bogoliubov description 
of a weakly interacting Bose gas. More generally, it is known \cite{gavoret}
 that this ratio $n_{0}/n$ is related to the zero momentum 
limit of the non-condensate particle distribution in the ground state, $n_k$, namely
\begin{equation}
{n_{k \rightarrow 0}}=\frac{n_0}n\frac{mc}{2\hbar k}
\label{n0}
\end{equation}
where $c$ is the sound velocity and $m$ is the mass of a $^4$He atom.
For the real liquid, we do not known how to obtain $n_{0}/n$ in terms of other measurable quantities. This ratio has been  
obtained indirectly \cite{sokol} from a measurement of the distribution of the non-condensed atoms. Numerically, using
 path integral Monte-Carlo \cite{ceperley} similar values have been obtained. An analytical expression has 
been proposed which is based on the assumption that the depletion of the condensate is related to the thermally excited rotons \cite{giorgini}.

In the present model, it is also possible to obtain the divergent part of the ground state occupation number $n_{k}=v(k)^2$ of
 the local modes. Using (\ref{uv}), it is given by
\begin{equation}
{n_{k \rightarrow 0}}=\frac{\Delta}{2\hbar kc}
\label{n0s}
\end{equation}
Since both (\ref{n0}) and (\ref{n0s}) describe the same behaviour, namely the condensation of the $^4$He atoms, 
then, by equating the two corresponding residues, we obtain the relation
\begin{equation}
\frac{n_0}n=\frac{\Delta}{m c^2}
\label{n0calc}
\end{equation} 
This relation predicts a direct and unexpected link between the condensate fraction $n_{0}/n$ and the characteristics of 
the energy spectrum at $T=0$, namely the sound velocity $c$ and the roton energy $\Delta$.
The behavior of both $n_{0}/n$ and ${\Delta}/{(m c^2)}$ measured as a function of the pressure are in agreement 
 with the relation (\ref{n0calc}) up to a factor of nearly two which we shall discuss later on.

Another relevant feature of (\ref{n0calc}) is that it allows to relate the
 quantity $n_{0}/n$ which is critical at the superfluid
 transition and therefore is proportional to the amplitude of the complex order parameter, to the roton energy $\Delta$ 
 which, in our model, is the modulus of the interaction matrix element at $k=0$ namely $\left| X(k=0) \right| =\Delta$.
A critical behavior of $\Delta$ was indeed recently observed \cite{montrooj} in high resolution neutron
 scattering experiments, and supported by Raman scattering experiments \cite{ohbayashi1}. This seems to indicate that 
the roton energy $\Delta$ (and therefore the interaction matrix element $X(k)$) 
is related to the superfluid order as it appears in (\ref{n0calc}).
 
\subsection{Condensation energy}

Another quantity which characterizes the condensate is the condensation energy $\Delta E_{G}$. It is obtained 
by rewriting $H_{loc}$ in (\ref{hloc}) in terms of the operators $\beta _k$ and $\beta _k^{\dagger }$, namely
\begin{equation} 
H_{loc}=\sum_kE(k)\beta _k^{\dagger }\beta _k+\sum_k\frac 12\big( 
E(k)-(E_0+X(k))\big)  \label{hlocb} 
\end{equation} 
 The second term in (\ref{hlocb}) is the shift in the ground state energy $\Delta E_G$. Using Eq.(\ref{ek}) we obtain 
\begin{equation} 
\Delta E_G=-\frac 1{8\Delta}\sum_k\left( E(k)-2\Delta \right)^2 
\label{degs} 
\end{equation} 
or
\begin{equation}
\Delta E_G=-\frac {V_{sp}}{32\pi^{2} \Delta}\int_{0}^{k_{max}}{k^2}\left( E(k)-2\Delta \right)^{2}{dk}
\label{degsi}
\end{equation}
where $k_{max}$ is the largest value of the momentum and corresponds to the termination point of the quasi-particle spectrum.
$V_{sp}$ is the specific volume per atom. Using the experimental \cite{cowley,pisto,donely} (fig.1) energy spectrum for
 $E(k)$, we obtain
\begin{eqnarray} 
\Delta E_{G}\simeq -5.0\pm 1.0K  (SVP) \nonumber \\
\Delta E_{G}\simeq -3.0\pm 1.0K  (P = 24 atm) 
\label{degtheo}
\end{eqnarray} 
for saturated vapor pressure and P=24atm respectively. 
In fig.4 we plot 
the integrand 
${k^2}\left( E(k)-2\Delta \right)^2$
in (\ref{degsi}). It shows that the main contribution to 
the reduction in the ground state energy comes from the roton part of the spectrum due to the 
larger phase space volume, compared with the phonon region ($k \rightarrow 0$). 

Both from experimental data \cite{sokol} and path integral Monte-Carlo calculations 
\cite{ceperley}, it is possible to deduce the values for the 
change in the kinetic zero point energy of the atoms between the normal phase at $T_{\lambda}$ and the superfluid phase 
in the limit $T \rightarrow 0$ 
\begin{eqnarray}
\Delta E_{kin}\simeq 2.5\pm 3.0K  (SVP) \nonumber \\
\Delta E_{kin}\simeq 1.3\pm 3.0K  (P = 24 atm)
\label{degexp}
\end{eqnarray}
for saturated vapor pressure and P=24atm respectively. 
 First, we notice that the large experimental uncertainty makes any quantitative comparison difficult. 
Moreover, the two quantities $\Delta E_{G}$ and $\Delta E_{kin}$ do not measure exactly the same property.
While $\Delta E_{G}$ measures the change in the ground state energy of the local modes between the non-interacting 
and interacting limits, both at T=0, $\Delta E_{kin}$ measures the change in kinetic energy per atom between 
two different temperatures. Since the variation of the volume of the superfluid between $T_{\lambda}$ and $T=0$ is
 small \cite{wilks}, it is usually accepted to atribute the change in the kinetic energy mainly to the condensation phenomenon.
 In our model this condensation energy is calculated per local mode, nevertheless we find that both quantities decrease
 for an increasing pressure.

\subsection{Effective mass}

Although the relations (\ref{n0calc}) and (\ref{degsi}) for the condensate fraction and the condensation energy give 
both the right order of magnitude and the expected bahaviour as a function of pressure, they are consistently 
larger than the experimental values by a factor of about 2. This factor may stem from the fact that our expressions
 are given as a function of the local modes while the experimental data is obtained per $^4$He atom, and there is no
 direct one-to-one correspondence between these two.
Therefore, the mass term which appears in (\ref{n0calc}) instead of being the bare mass $m$ of a $^4$He atom is the
 effective mass $m_{eff}$ of the bare local mode. From the above comparison between $n_{0}/n$ and $\Delta E_{G}$
 with experiments we obtain that $m_{eff} \simeq 2m$. Although such a relation can be obtained only from a 
microscopic calculation beyond our phenomenological model, it is interesting to notice that a 
similar result was obtained numerically using path integral Monte-Carlo calculations \cite{ceperley}. There, the effective mass $m^*$ 
of a tagged $^4$He atom which does not participate in the Bose permutations was found to be $m^{*} \simeq 2m$.
 Since such a tagged atom is a local node or defect in the superfluid, it corresponds to the {\em bare} local mode
 we consider in the Hamiltonian $H_{loc}$. As we saw in (\ref{uv}) and (\ref{psi0}), the bare local mode of energy $E_0$
 appears for the non-interacting case ($X(k)=0$ and $v(k)=0$). Then, it does not contribute to the coherent (superfluid) 
ground state, i.e. behaves as a local node of the order parameter. This correspondence may justify taking 
$m_{eff} \simeq 2m$ bringing both (\ref{n0calc}) and (\ref{degsi}) into agreement with the experimental data
represented in fig.5, where the experimental values \cite{sokol} for $n_{0}/n$,
 $\Delta$ and $c$ are obtained independently \cite{wilks}.

\section{Hybridization of localized and density modes} 

In the previous sections we described the superfluid as resulting from the condensation of a set of localized 
and interacting modes in the strong coupling limit. This description, as we emphasized, is very different from 
the Bogoliubov weak coupling limit.
Since it is aimed to describe the full energy spectrum, we would like to relate it to the Feynman 
variational scheme. In this section we shall show that in addition to the phonon-roton branch of excitations 
around the roton momentum, there is another branch of excitations.
 We shall discuss the properties of this excitation in the light of experiments on thermal nucleation of vortices 
\cite{varoquaux} and on large momentum neutron scattering \cite{fak}.
 To that purpose, we propose in this section a description equivalent to the previous interacting local mode problem, where
 the interaction between these modes results from the exchange of virtual phonons. This is very much in the
 same spirit of Hopfield \cite{hopfield} and Anderson considering the exciton problem in dielectric media.
There localized excitons, taken as dipoles, interact through the exchange of photons. For the superfluid, 
the local modes play the role of the excitons, and interact through the exchange of Feynman phonons. 
This picture is only valid in the range of large enough momentum beyond the linear part around $k \rightarrow 0$,
 which corresponds to density excitations (i.e. Feynman phonons). It is only in this regime that the two types of excitations
 represent independent degrees of freedom of the system. The Hamiltonian describing Feynman density excitations is
\begin{equation} 
H_0=\sum_k\varepsilon (k)a_k^{\dagger }a_k  \label{h0} 
\end{equation} 
 where $a_k$ are Bose operators and $\varepsilon (k)$ is the 
Feynman spectrum \cite{feynman} given by the relation (\ref{feynspec}), expressed in terms of the static 
structure factor $S(k)$.
The Hamiltonian describing the coupling between these phonons and the localized modes 
is ~\cite{hopfield}  
\begin{eqnarray} 
H_c &=&{\sum_k}\left( \lambda (k,{E_0}){b_k}+\mu (k,{E_0}){a_k}%
\right) ({{a_k}^{\dagger }}+{a_{-k}})  \label{hc} \\ 
&+& h.c.  \nonumber 
\end{eqnarray} 
and the total Hamiltonian is
\begin{equation}
H=H_{0}+H_{loc}^{0}+H_{c}
\label{htot}
\end{equation} 
Since $H$ and $H_{loc}$ provide two equivalent descriptions of the same problem, the functions 
$\lambda $ and $\mu $ 
are related to $X(k)$. If we now assume that the localized modes are local dipolar excitations, then there is a specific
 relation \cite{hopfield,gov1} between these functions, given by
$\lambda (k) = i {E_0} \left(-{3X(k) \over {2 \varepsilon (k)}} \right)^{1 \over 2}$
and $\mu(k) = - {E_0} {{3 X(k)} \over {2 \varepsilon (k)}}$.
As we shall discuss later on, this dipolar approximation is valid only for large momentum, i.e. in the roton region. 
The two modes $a_k$ and $b_k$ are independent degrees of freedom, so that their commutator is zero.
The total Hamiltonian $H$ can be diagonalized using the canonical transformation \cite{hopfield}  
\begin{eqnarray} 
\alpha _k &=&A(k)a_k+B(k)b_k+C(k){a^{\dagger }}_{-k}+D(k){b^{\dagger }}_{-k} 
\label{alfa} \\ 
{\tilde \alpha }_k &=&B(k)a_k+A(k)b_k+D(k){a^{\dagger }}_{-k}+C(k){b^{\dagger }}_{-k}  
\nonumber 
\end{eqnarray} 
which describe respectively the lower and upper branch of the energy spectrum. 
The functions $A(k),B(k),C(k),D(k)$ can be 
written down explicitly \cite{hopfield}. We point out that the roles of the 
phonons and localized-modes are interchanged between the two branches, as 
can be seen by writing the ground-states of both branches:  
\begin{eqnarray} 
\left| 0_1\right\rangle  &=&\prod_k\exp \left( \frac {C(k)}{A(k)}{a^{\dagger }}_k{%
a^{\dagger }}_{-k}\right) \exp \left( \frac {D(k)}{B(k)}b{^{\dagger }}_kb{^{\dagger }}%
_{-k}\right)   \nonumber \\ 
\left| 0_2\right\rangle  &=&\prod_k\exp \left( \frac {D(k)}{B(k)}{a^{\dagger }}_k{%
a^{\dagger }}_{-k}\right) \exp \left( \frac {C(k)}{A(k)}b{^{\dagger }}_kb{^{\dagger }}%
_{-k}\right)   \label{psi0hyb} 
\end{eqnarray} 
The corresponding dispersion relation is  
\begin{equation} 
\left[{E^2}(k)-{\varepsilon ^2}(k)\right]\left[{E^2}(k)-{E_{0}^2}\right]=-6X(k)E_{0}{E^2}(k)
\label{ehop} 
\end{equation}  
We first notice that for a vanishing coupling $X(k) = 0$, we obtain, as expected, the two solutions 
$E_{1}(k)=\varepsilon(k)$ and $E_{2}=E_{0}$, describing respectively pure density and non interacting localized modes. 
A non-zero coupling hybridizes these two sets of excitations. 
Since $X(k)$ and $E(k)$ are also related by the relation (\ref{ek}) 
\begin{equation}
{E^2}(k)-{E_{0}^2}=2X(k)E_{0}
\label{xxk}
\end{equation}
 using the relations 
(\ref{xxk}) and (\ref{ehop}), we obtain a two branch spectrum of dispersions
\begin{eqnarray}
E_{1}(k)&=&{\frac 12}\varepsilon (k)={\frac{{\hbar^2}{k^2} }{{4m S(k)}}} \nonumber \\
E_{2}&=&2{E_0}=4\Delta
\label{branch}
\end{eqnarray}
These expressions are written in terms of $\varepsilon (k)$ and $\Delta$, but they depend also explicitely
 on $X(k)$ through (\ref{ek}), which is a consequence of the dipolar approximation.
As a result of the hybridization, the energy spectrum $E_{1}(k)$ of the density fluctuations
is shifted by a factor 2 towards lower energies relatively to the Feynman
ansatz. There is in addition a new branch $E_{2}$ of localized excitations at a constant
energy which is twice the value of the bare local mode.
It is of some interest to notice that it is possible to express the factors 
$\left| \frac {C(k)}{A(k)} \right|$ and $\left| \frac {D(k)}{B(k)} \right|$
 appearing in the coherent ground-state wavefunctions of the two branches
(\ref{psi0hyb}) in terms of the energies, namely
\begin{eqnarray} 
\left| \frac {C(k)}{A(k)}\right| &=&\left| \frac{\varepsilon (k)-E(k)}{\varepsilon (k)+E(k)}%
\right| \nonumber \\
\left| \frac {D(k)}{B(k)}\right| &=&\left| \frac{{E_0-E(k)}}{{E_0+E(k)}}\right|
\label{hyboccu} 
\end{eqnarray} 
For the special value $k=k_{rot}$ which corresponds to the roton minimum, they are both equal, 
$\left| C/A\right|(k_{rot})=\left| D/B\right|(k_{rot})=1/3$, a consequence of the relation (\ref{branch}).
 The roton excitation then appears as involving the maximum mixing of localized and density modes 
in equal proportions.

We want now to compare these results with the experimental data \cite{pisto,cowley,donely} obtained
for the energy spectrum $E(k)$ (fig.6) and for the structure factor $S(k)$ \cite{sven,hen,hall} at 
two different pressures (fig.7).
Around the roton minimum, and over quite a
large range of momentum, the relation (\ref{branch}) is in good agreement with the
experimental results and gives a spectrum which is reduced from the Feynman result by a factor of nearly two.
 It is interesting to recall that Feynman himself in his original paper \cite{feynman} noticed the factor 2
discrepancy between his ansatz and the experimental results around
the roton minimum. Here we deduce it from a closed analytical solution. This interpretation of the roton
excitation as resulting from the dipolar hybridization of two separate
excitations is to be compared with the approach of Glyde and Griffin ~\cite
{Glyde and Griffin,griffin} which uses a dielectric formalism to
describe hybridized phonons and free-particle excitations.
The main discrepancy is obtained in the
low momentum region ($k \leq 1 {\AA}^{-1}$), i.e. below the maxon momentum.
As $k \rightarrow 0$ the experimental data is close to the Feynman result
i.e. twice the value we obtain in (\ref{branch}). This regime is beyond the
dipolar approximation we consider and sets the limit of validity of this approach. To be able to describe 
the linear phonon regime from the present approach, we should include multipolar terms.  
 
\section{High energy excitation branch}

\subsection{Vortex loops}

The relation (\ref{branch}) gives a second branch of excitations at the constant
 energy $E_{2}=4\Delta $.
It describes a localized (dispersionless) excitation of energy which is twice the
bare vortex core (local mode) energy.
The intuitive physical picture of this mode is therefore
of an excitation made out of two
vortex-core excitations. Such an excitation in three
dimensions may be viewed as a microscopic vortex-loop, or a localized defect.
The radius $R$ of a vortex-loop can be estimated using a Feynman-type formula for
the energy of the associated circulating current ~\cite{nozieres}
\begin{equation}
E_{vortex}=2{\pi ^2}\rho {\frac{{{\hbar ^2}R}}{{m^2}}}ln\left( {\frac Ra}%
\right)
\label{evort}
\end{equation}
where at $T=0$, we take the density of the superfluid ${\rho _s}$ to be the total
density $\rho$
and $a$ is the core size equal to the atomic radius namely $a\simeq 1.4\AA $%
 \cite{core}. The radius $R$ obtained from (\ref{evort}) and which corresponds to
$E_{vortex}=4\Delta =34.4K$
is $R\simeq 5.1\AA $.

Experimental support for the interpretation of the hybridized
localized state at $E=4\Delta $ as an intrinsic excitation of the fluid is
provided by critical velocity experiments~\cite{varoquaux}. In
phase-slippage studies of the critical velocity through an orifice, the
critical velocity is driven by the thermal nucleation of vortex-loops. The
corresponding activation energy $E_v$ is determined by the nucleation rate 
$\Gamma $
given by the Arrhenius law $\Gamma ={\Gamma _0}exp\left( {%
\frac{{-{E_v}}}{{{k_B}T}}}\right) $ and turns out to be ~\cite{varoquaux} ${E_v%
}\simeq 33\pm 5K$ a value indeed very close to $4\Delta =34.4K$.
Moreover, the upper critical velocity $v_c$ may be estimated to be
 the velocity of the vortex-loop itself ~\cite{nozieres} i.e. ${v_c}={%
\frac{{\hbar }}{{2mR}}}ln\left( {\frac Ra}\right) \simeq 20m/s$, a value
close to the largest measured critical velocity ~\cite{varoquaux}.

Finally, recent experiments \cite{maris} on cavitation in superfluid $^4$He
are analyzed in terms of thermal a nucleation process. The activation
energy found is of the order of $4 \Delta$, indicating a possible
connection between a microscopic vortex-loop and the thermal nucleation
of cavitation.

We also point out that this interpretation of the vortex-loop branch as distinct from the
phonon-roton branch is further supported experimentally by ion velocity
measurements \cite{mclintock} and theoretically using hydrodynamic
methods \cite{jones}.

\subsection{High energy neutron scattering}

Recent high resolution neutron scattering data show that at high energy
and momentum transfer, compared with the phonon-roton branch, the
position of the scattering peak \cite{fak} is consistently at a lower
energy than the free-recoil value.
At such high energies the atoms of the
liquid behave freely \cite{glyde}, therefore with a quadratic dispersion
$E_{free}(k)=\hbar ^2k^2/2m$.
There is a marked broadening of the width of
this peak where the free-recoil spectrum crosses the second branch of energy
$4\Delta$ at
momentum $k_c\simeq 2.3\AA ^{-1}$.
This data can be understood as resulting from the superposition
of the free-recoil spectrum and the process of vortex-loops creation. Such
a possibility for explaining the increased width of the dispersion curve
using vortex-loop creation was suggested by Cowley and Woods \cite{woods}. A
recoiling atom can create vortex-loops and loose the corresponding energy and
momentum, so that its remaining energy is
\begin{equation}
E_{n}(k)=4n\Delta +\frac{\hbar ^2\left( k-k_{c,n}\right) ^2}{2m}  \label{esplit}
\end{equation}
where $n$ is the number of emitted vortex-loops and $k_{c,n}$ is given
by the equality of the free-recoil energy and the $n$th vortex-loop energy, so
that
$k_{c,n}\simeq 2.3\sqrt{n}\AA ^{-1}$. We show the dispersion for the
first two emitted vortex-loops in fig.8.
 Then, by substracting from the experimental neutron-scattering data a Lorentzian
\cite{griffin} centered around the free-recoil energy and fitted to the
highest energy tail, we find an
additional scattering peak around the energy given by (\ref{esplit}).
In fig.9 the experimental results \cite{fak} are plotted, which
correspond to $k$'s such that $n=1$ or $2$.
At higher energy transfer
the free-recoil spectrum can be split again at twice $4\Delta $ and so on, as is
indicated in fig.9d. There is a series of such splittings each
time the recoiling atom has an energy which is an integer multiple of $4\Delta $.
In fig.10, the experimental results \cite{martel}
for the high momentum transfer $k=10\AA ^{-1}$ are shown. At this momentum the 
free-recoil energy is more than
17 times the vortex-loop energy of $4\Delta $,
and we find that
the extra scattering is well described as a
'band' of splitted levels below the free-recoil spectrum.

\section{Conclusion}

In this paper, we have presented a phenomenological strong coupling theory for 
the superfluid phase of $^4He$ in
order to describe features that correspond to excitations of energies beyond the
 phonon part of the spectrum. This
approach is based on the assumption that at short length scales, there exists a set
of interacting
localized excitations. This is very much in the spirit of the assumption of two-
level tunneling states in
dielectric glasses. These excitations are described by an effective Hamiltonian
which depends on the strength
$X(k)$ of the interaction between these modes. We used experimental results
obtained in neutron scattering in order
deduce the behaviour of $X(k)$. This approach gives a hand to calculate analytical
expressions for various physical
quantities like the scattering cross-section $Z(k)$ for single particle excitations,
and the shift in the ground state
energy. In addition, it allowed us to derive a relation between the condensate 
fraction and characteristics of the
energy spectrum namely the sound velocity and the roton energy $\Delta$. This 
gives a way to measure directly the
condensate fraction. Moreover, it establishes the critical nature of $\Delta$ 
and its relation to superfluidity in
close connexion with recent experimental results.
Then, we derived an alternative description of the set of localized modes which
allowed us to predict the existence
of a new branch of excitations at energy $4 \Delta$ beyond the phonon-roton branch.
This branch which has not yet been
observed directly shows up indirectly and allows to interpret phase slippage exper
iments and to get a hint for the puzzling
broadening observed at high energy and momentum transfer neutron scattering.
Since we interpret this excitation as the smallest vortex-loop (or local
defect) of the superfluid, our model offers a way of describing both the
phonon-roton and the localized vortex-loops within one scheme
\cite{putterman}.

{\bf Acknowledgement} It is our pleasure to acknowledge the very kind support of both 
the LPTMS (Bat.100) and the department of Solid state physics (Bat.510) at the university 
of Paris Sud (Orsay), and G. Montambaux and A. Comtet for their interest.

\newpage
 
\begin{figure}
\input epsf
\centerline{ \epsfysize 11.0cm
\epsfbox{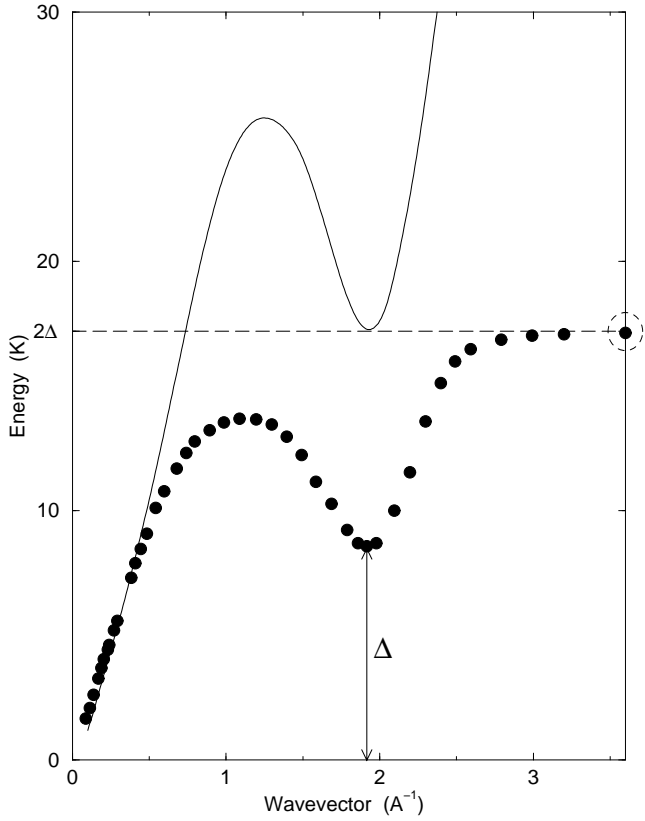}}
\caption{The experimental energy spectrum at saturation vapor
pressure [3,9,32] compared with the Feynman phonon spectrum (Eq.1) based on
the experimental static structure factor $S(k)$ [33-35]. Also marked is
the roton energy $\Delta$ and the termination of the quasi-particle
branch at energy $2 \Delta$ (circled point).}
\end{figure}
\begin{figure}
\input epsf
\centerline{ \epsfysize 11.0cm
\epsfbox{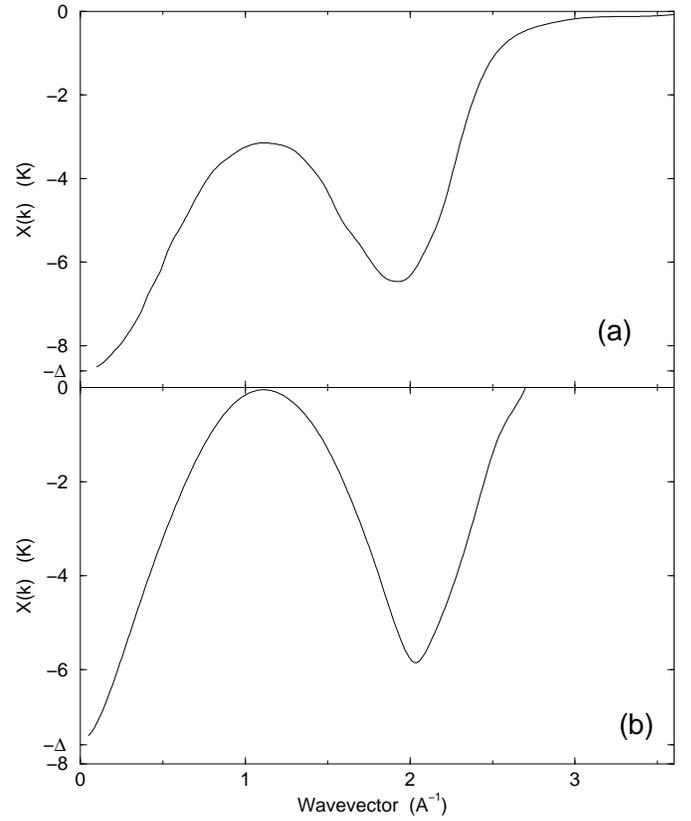}}
\caption{The interaction matrix $X(k)$ from Eq.(4), based on the experimental energy
spectrum at (a) saturation vapor
pressure [3,9,32] and (b) P=24 atm [3]. 
The bare energy of the local mode is taken to be $E_0=2\Delta$.
Notice that $X(k)$ tends to zero at the termination point of the
spectrum at high momentum.}
\end{figure}
\begin{figure}
\input epsf
\centerline{ \epsfysize 11.0cm
\epsfbox{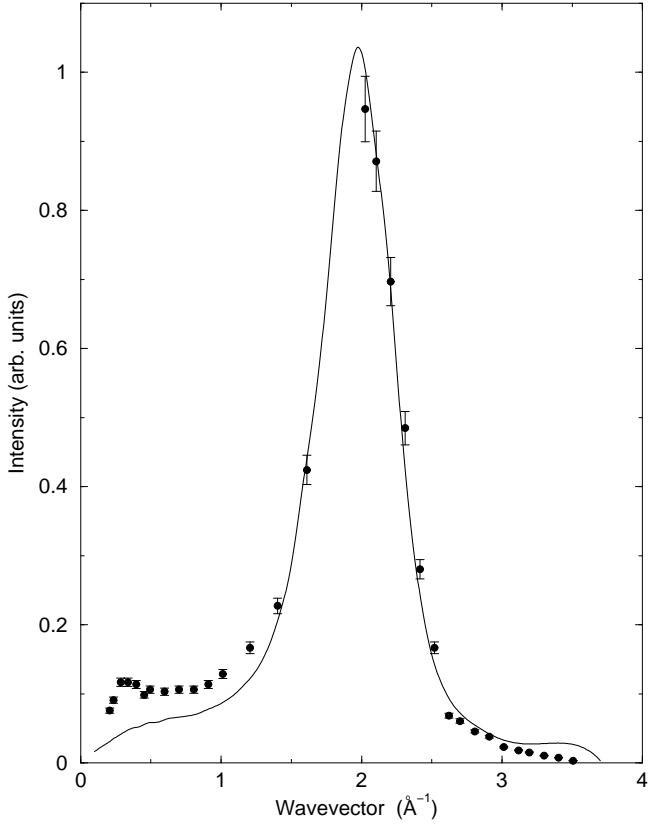}}
\caption{Comparison between the experimental scattering cross-section
[3] $Z(k)$ of single quasi-particle excitations (points) at 1.1K and
the theoretical curve (solid line), obtained using
the experimental data for
the energy spectrum at saturated vapor pressure $E(k)$ [3,9,32] in Eq.(12).}
\end{figure}
\begin{figure}
\input epsf
\centerline{ \epsfysize 11.0cm
\epsfbox{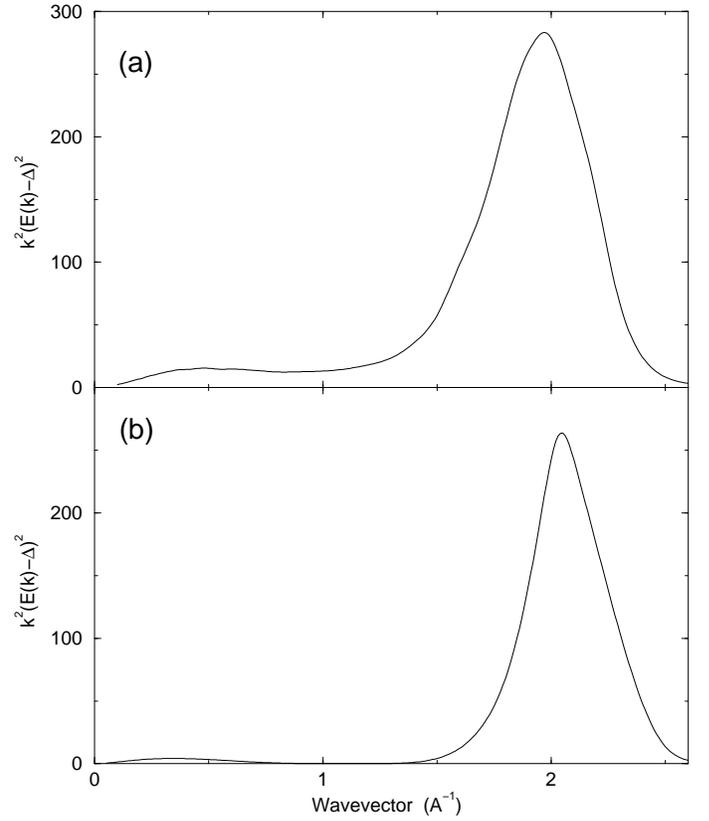}}
\caption{The integrand 
${k^2}\left( E(k)-2\Delta \right)^2$
appearing in the
calculation of the reduction in the ground state energy $\Delta E_G$ in Eq.(18).
(a),(b) are at saturation vapor pressure and P=24atm respectively.
Notice the large contributions around the roton momentum.}
\end{figure}
\begin{figure}
\input epsf
\centerline{ \epsfysize 11.0cm
\epsfbox{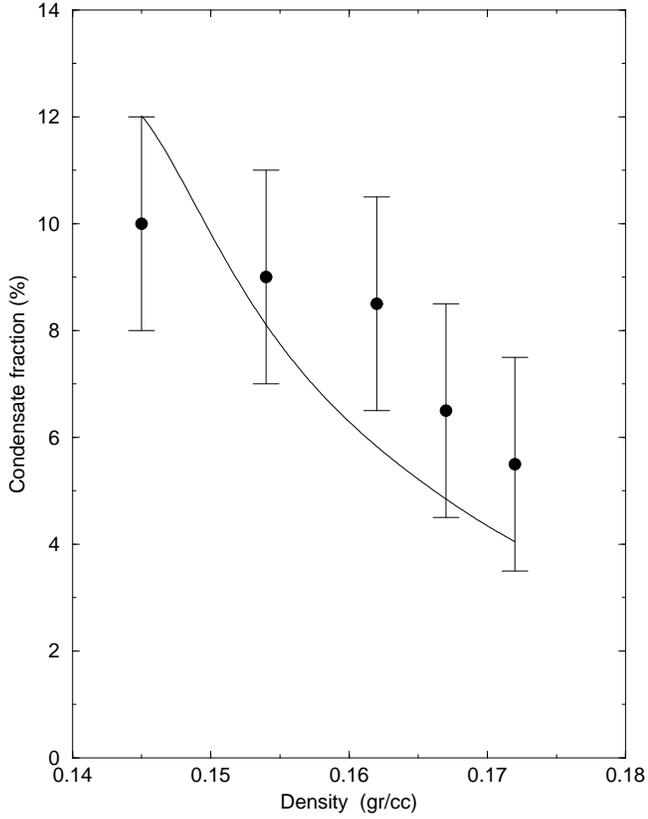}}
\caption{Comparison between the experimental condensate fraction at
different densities (solid circles) [23] and Eq.(15), using an effective
mass for the local modes of $m_{eff}=2.3 m$ (solid line).}
\end{figure}
\begin{figure}
\input epsf
\centerline{ \epsfysize 11.0cm
\epsfbox{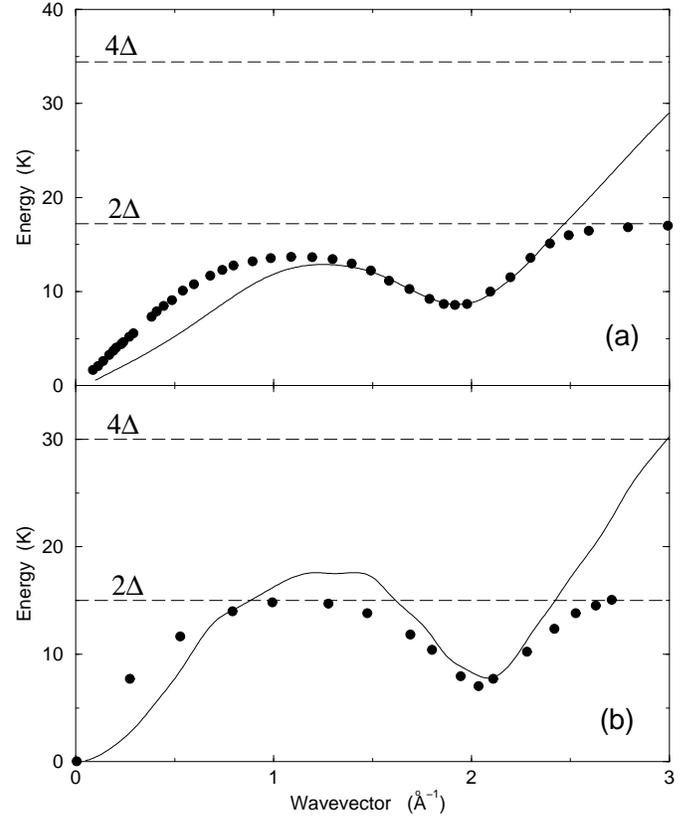}}
\caption{Comparison between the experimental energy spectrum
[3,9,32] (points) and
the energy $E_{1}(k)$ in Eq.(28) (solid line), where the structure factor
$S(k)$ is
obtained from independent measurements [33-35]. (a) and (b)
correspond respectively to
the saturation vapor pressure and to P=24 atm. The dashed line at
energy $E_{2}=4\Delta$ indicates the position of the localized branch of
excitations (vortex-loop).}
\end{figure}
\begin{figure}
\input epsf
\centerline{ \epsfysize 11.0cm
\epsfbox{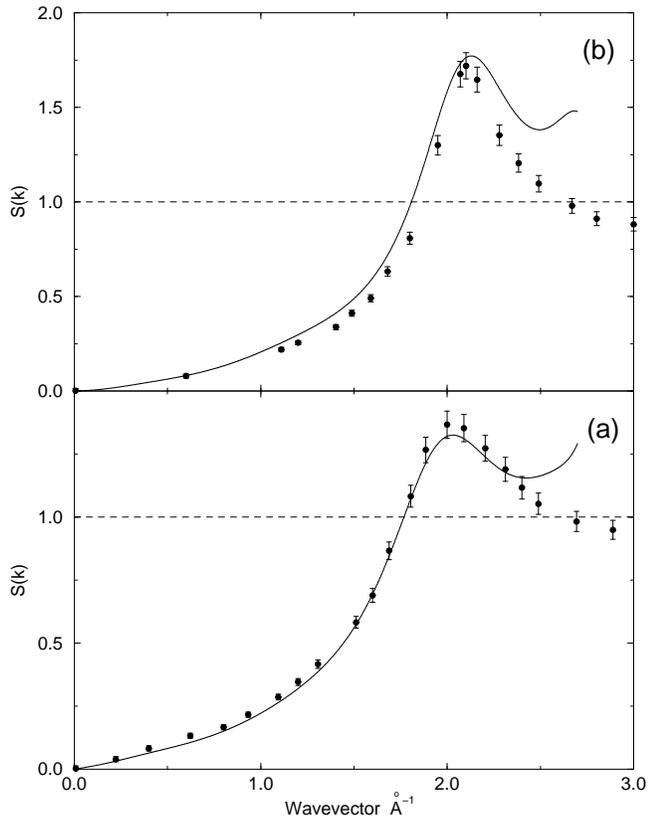}}
\caption{Comparison between the experimental structure factor
$S(k)$ [33-35] (solid circles) and expression Eq.(28) (solid line) for the
same two pressures as in Fig.6, where the energy $E(k)$ is obtained
from independent measurements [3,9,32].}
\end{figure}
\begin{figure}
\input epsf
\centerline{ \epsfysize 11.0cm
\epsfbox{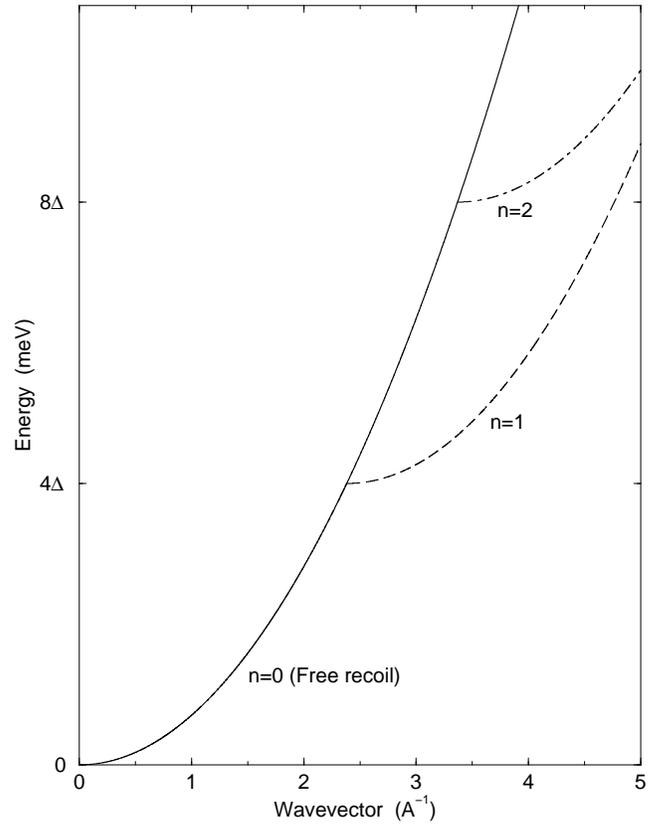}}
\caption{The dispersion relations of the free recoil ($n=0$) and the
lowest two
levels of recoiling atoms that correspond to vortex-loops emission ($n=1,2$) (Eq.31).}
\end{figure}
\begin{figure}
\input epsf
\centerline{ \epsfysize 11.0cm
\epsfbox{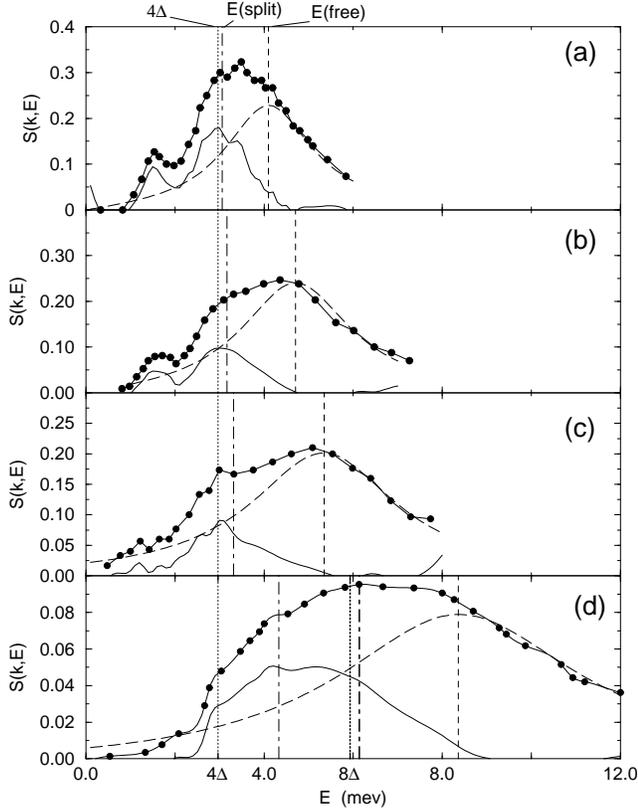}}
\caption{Comparison between the experimental scattering
profiles [29] (solid circles) and the free-recoil Lorentzian
(heavy dashed-line) at different momenta: (a) $k=2.8 \AA ^{-1}$, (b)
$k=3.0 \AA^{-1}$, (c) $k=3.2 \AA ^{-1}$, (d)
$k=4.0 \AA ^{-1}$. The remaining extra scattering is the heavy
solid-line. The
vortex-loop energy ($4 \Delta$) is indicated by the vertical
dotted-line, the free-recoil
energy by the vertical long-dashed-line, and the $n=1$ splitted energy
Eq.(31) by the vertical dashed-dot line.
At relatively low momenta (a,b)
there is still a noticable phonon-roton peak around the $2\Delta$
energy.
In (d) the heavy vertical lines indicate the $n=2$ vortex-loop
emission by the recoiling atom.}
\end{figure}
\begin{figure}
\input epsf
\centerline{ \epsfysize 11.0cm
\epsfbox{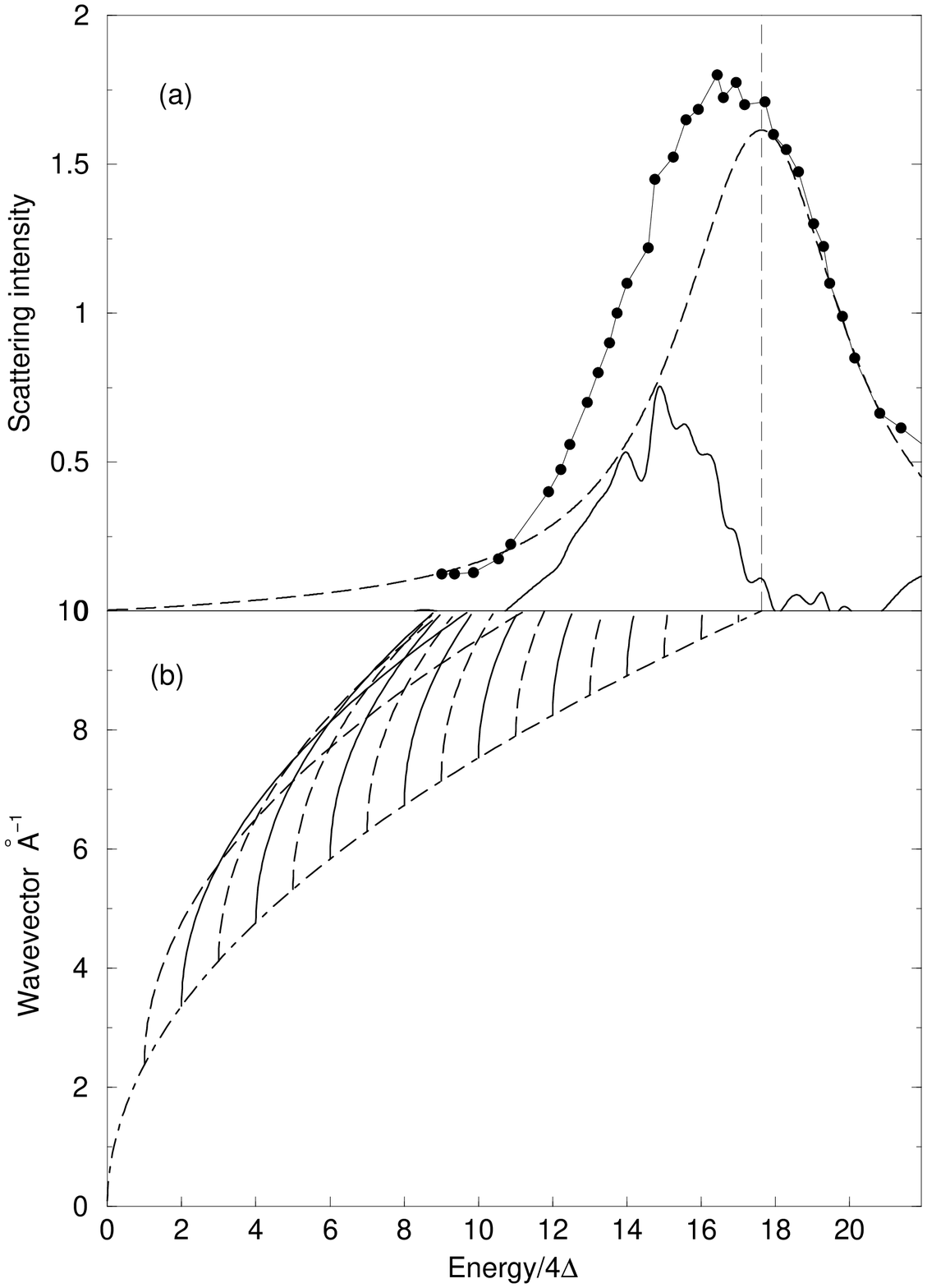}}
\caption{(a) Experimental scattering profile [44] at
$k=10.0 \AA ^{-1}$ (solid circles). The free-recoil Lorentzian (heavy
dashes-line) and extra
scattering (heavy solid-line) are plotted. The free-recoil energy is
indicated by the long-dashed vertical line. (b) The spectrum of the
free-recoil
(dash-dot) and the 17 splittings due to
vortex-loop excitations (Eq. 31) (alternating solid and dashed lines).}
\end{figure}


\begin{references}
\bibitem{note} This choice of the bare energy $E_0$ could also be understood in the following way.
 Consider an impurity atom of $^3 He$ in superfluid $^4He$. This impurity is a local node of the 
order parameter and as such should have an excited state of energy $E_0$. This energy have been measured 
in a similar situation namely in bcc $^4 He$, for which the NMR data of a $^3 He$ impurity atom has been used 
to obtain the value of the bare local mode energy \cite{bcc}. Both for saturated vapor pressure 
and for P=19 atm and for $T \rightarrow  T_\lambda$, there was measured
\cite{garwin,ptukha,opfer} a thermally activated component to the 
 $^3 He$ spin relaxation with energy close to $E_0 = 2 \Delta $.
\end{references}
\end{document}